\documentclass[aps,prd,twocolumn,showpacs,preprintnumbers,nofootinbib,amsmath,amssymb]{revtex4} 
\usepackage{graphicx}
\usepackage{bm}
\usepackage{amsmath}
\usepackage[normalem]{ulem}
\usepackage{multirow}

\newcommand{\beq}{\begin{eqnarray}}
\newcommand{\eeq}{\end{eqnarray}}

\newcommand{\tr}{\ensuremath{\mathrm{Tr}}}

\def\spose#1{\hbox to 0pt{#1\hss}}
\def\ltapprox{\mathrel{\spose{\lower 3pt\hbox{$\mathchar"218$}}
 \raise 2.0pt\hbox{$\mathchar"13C$}}}

\begin{document}

\title{
Confining and chiral properties of QCD 
in extremely strong magnetic fields
}

\author{Massimo D'Elia}
\email{massimo.delia@unipi.it}
\affiliation{
Dipartimento di Fisica dell'Universit\`a
di Pisa and INFN - Sezione di Pisa,\\ Largo Pontecorvo 3, I-56127 Pisa, Italy}

\author{Lorenzo Maio}
\email{lorenzo.maio@phd.unipi.it}
\affiliation{
Dipartimento di Fisica dell'Universit\`a
di Pisa and INFN - Sezione di Pisa,\\ Largo Pontecorvo 3, I-56127 Pisa, Italy}

\author{Francesco Sanfilippo}
\email{francesco.sanfilippo@infn.it}
\affiliation{INFN - Sezione di Roma Tre,\\ Via della Vasca Navale 84, I-00146 Rome, Italy}

\author{Alfredo Stanzione}
\email{a.stanzione1@studenti.unipi.it}
\affiliation{
Dipartimento di Fisica dell'Universit\`a
di Pisa and INFN - Sezione di Pisa,\\ Largo Pontecorvo 3, I-56127 Pisa, Italy}

\date{\today}

\begin{abstract}
We investigate, by numerical lattice simulations, 
the static quark-antiquark potential, the flux tube 
properties and the chiral condensate 
for $N_f = 2+1$ QCD with physical quark masses 
in the presence of strong magnetic fields, going 
up to $eB = 9$~GeV$^2$, with continuum extrapolated results.
The string tension for 
quark-antiquark separations longitudinal to the magnetic field
is suppressed by one order of magnitude at the largest explored
magnetic field with respect to its value at zero 
magnetic background, but is still non-vanishing; in the transverse direction, instead,
the string tension is enhanced but seems to reach a saturation 
at around 50~\% of its value at $B = 0$. The flux tube 
shows a consistent suppression/enhancement of the overall amplitude, 
with mild modifications of its profile.
Finally, we observe magnetic catalysis in the whole range 
of explored fields with a behavior compatible with a lowest Landau
level approximation, in particular with a linear dependence
of the chiral condensate on $B$ which is in agreement, within
errors, with that already observed for $eB \sim 1$~GeV$^2$.
\end{abstract}

\maketitle

\section{Introduction}
\label{intro}

In the recent past, various analytic and numerical studies 
have uncovered a plenty of interesting new phenomena 
regarding the non-perturbative properties of strong 
interactions in the presence of a magnetic background field~\cite{Yamamoto:2021oys,Ding:2020hxw,Hofmann:2020ism,Cao:2019azh,lecnotmag,anisotropic,chernodub,musak,elze1,elze2,mueller,galilo,simonov2,KojoSu1,KojoSu2,watson,andersen,ozaki,kamikado,mueller2,demusa,DEN,Ilgenfritz:2012fw,reg2,EB,kovacs,Ilgenfritz:2013ara,DElia:2018xwo,catalreview,fukuhida,Bornyakov:2013eya,Chao:2013qpa,Fraga:2013ova,Yu:2014sla,Ferreira:2014kpa,Farias:2014eca,Ruggieri:2014bqa,strongmag0,strongmag1,tusso,Bonati:2017uvz,DElia:2015eey,Bali:2011qj,Bali:2012cd,rhomass1,rhomass2,rhomass3,rhomass4,simonov,Kojo:2021gvm,Hattori:2019ijy,Machado1,Machado2,Alford,Filip,Dudal:2014jfa,Cho:2014loa,Bonati:2015dka,Suzuki:2016kcs,Finazzo:2016mhm,Yoshida:2016xgm,Suzuki:2016fof,Iwasaki:2018pby,Iwasaki:2018czv,Khan:2021syq,Iwasaki:2021nrz,Zhou:2020ssi}.
Some of these phenomena might be of direct phenomenological relevance 
for heavy ion experiments~\cite{hi1,hi2,hi3,hi4,tuchin} or astrophysics~\cite{magnetars,vacha,grarub}, some of them are
more speculative but nevertheless interesting.
Among these phenomena, a direct impact on the QCD vacuum properties,
in particular those regarding the pure gauge sector, is particularly striking,
since gluons are not electrically charged, and might be the signal
of a stronger impact of the magnetic field on the QCD phase
structure.

In Refs.~\cite{strongmag0,strongmag1}, a direct effect on the 
static quark-antiquark potential has been unveiled, consisting
mostly of a suppression of the string tension for quark-antiquark
separations parallel to the magnetic background, and of an enhancement
for transverse separations; such findings have been confirmed
in Ref.~\cite{tusso} by a direct 
investigation of the color flux tube properties and 
can be interpreted within various model computations~\cite{anisotropic, galilo, Giataganas:2012zy,
Ferrer:2014qka, Rougemont:2014efa, Chernodub:2014uua, Miransky:2015ava,
Simonov:2015yka, Endrodi:2015oba, Schafer:2015wja, Dudal:2016joz,
Hasan:2017fmf, Giataganas:2018uuw,  Andreichikov:2018wrc}.

In particular, the conclusions of Ref.~\cite{strongmag1}
pointed to the possible presence of a critical magnetic field
$eB \gtrsim 4$~GeV$^2$, above which the longitudinal string 
tension would vanish, resulting in a different and yet unknown 
phase of strongly interacting matter.
Such conclusions, however, were not based on direct simulations
performed at such large values of the magnetic background, but
just on the extrapolation of results obtained in a smaller magnetic
field range.

Given the new and interesting predicted phenomena, a direct investigation
is of utmost importance. As we will better explain
in the following Sections, the main difficulty in studying
large magnetic backgrounds by lattice simulations
is that the ultraviolet (UV) cut-off must be tuned correspondingly
in order to keep discretization errors under control and allow 
for a reliable continuum extrapolation. In this study we will
investigate $N_f = 2+1$ QCD with physical quark masses and lattice
spacings down to $a \simeq 0.057$~fm, which is around half the 
finest spacing explored in Ref.~\cite{strongmag1}, with a 
similar discretization based on stout-improved staggered fermions. 
That 
will allow us to obtain continuum extrapolated results
for $eB$ up to $\sim 10$~GeV$^2$, which is enough to confirm 
or update the prediction of Ref.~\cite{strongmag1}.
In addition to that, we will consider the chiral properties
of the theory, in particular the chiral condensate, to investigate
if magnetic catalysis is still at work.

The paper is organized as follows. 
In Section~\ref{methods} we provide more details regarding the 
adopted discretization of $N_f = 2+1$ QCD in the presence
of a magnetic background, as well as about the lattice observable
used to extract the potential and the 
chromoelectric field between the static quark-antiquark pair.
In Section~\ref{results}, after some preliminary details regarding
our numerical simulations, we illustrate our results for
the chiral condensate, the static potential and the color flux tube.
Finally, in Section~\ref{conclusions}, we summarize our conclusions.

\section{Numerical Methods}
\label{methods}

We consider $N_f = 2+1$ QCD in the presence of a uniform and 
constant, external magnetic field, discretized in terms of 
the tree-level improved Symanzik action~\cite{weisz,curci} 
for the gauge sector, and of rooted staggered fermions with stout 
improvement~\cite{kogut-susskind,morning} for the fermionic sector. 
The resulting partition function is
\begin{equation}
  Z=\int{[DU]}\,e^{-S_{YM}}\prod_{f=u,d,s} \det{(D_{st}^f)}^\frac{1}{4},
\end{equation}
where $[DU]$ is the $SU(3)$ group invariant integration measure on the link variables, $f$ the flavor index,
\begin{equation}
  S_{YM}=-\frac{\beta}{3} \, \sum_{\substack{i \\ \mu\ne\nu}}\left(\frac{5}{6}W^{1\times1}_{i,\mu\nu}-\frac{1}{12}W^{1\times2}_{i,\mu\nu}\right)
\end{equation}
is the lattice gauge action, and
\begin{equation}\label{dirac_operator}
  \left(D^f_{st}\right)_{ij}=am_f\delta_{i,j}+\sum_{\nu=1}^{4}\frac{\eta_{i;\nu}}{2}\left(U^{(2)}_{i;\nu}\delta_{i,j-\hat{\nu}} - U^{(2)\dagger}_{i-\hat{\nu};\nu}\delta_{i,j+\hat{\nu}}\right)
\end{equation}
is the discretized Dirac operator. There, $i$ labels lattice sites and $\mu$ the direction, while $\beta$ is the inverse gauge coupling and $a$ the lattice spacing. The $W^{1\times \cdot}_{i,\mu\nu}$s are the real parts of the trace of the links products along the $1\times1$ and $1\times2$ rectangular closed path, respectively. The $\eta_{i;\nu}$ are the staggered quark phases, and $U^{(2)}_{i;\nu}$ is the two times stout smeared link (with isotropic smearing parameter $\rho=0.15$).

An external electromagnetic (e.m.) field is added by minimal substitution in the covariant derivative
\begin{equation}\label{covariant_der}
  \partial_\mu + i g_0 A_\mu^a(x) T^a \to  \partial_\mu + i g_0 A_\mu^a(x) T^a + i q_f A_\mu(x),
\end{equation}
where $A_\mu^a(x)$ are the gluon fields, $T^a$ the $SU(3)$ generators, $A_\mu(x)$ the abelian four-potential, and $g_0$ and $q_f$ are respectively the bare strong coupling constant and the quark electric charge. We consider for simplicity 
a uniform magnetic field $\vec{B}$ in the $\hat{z}$ direction, 
a possible gauge choice is then:
\begin{equation}\label{four_potential}
  A_t=A_x= A_z= 0, \qquad A_y(x) = Bx.
\end{equation}
That can be discretized on a periodic toroidal lattice
in terms of $U(1)$ link variables as follows
\begin{equation}
  u_{i;y}^f=e^{i a^2 q_f B \,i_x}, \qquad {u_{i;x}^f\vert}_{i_x=L_x}=e^{-ia^2q_fL_xBi_y},
\end{equation}
where $L_x$ is the lattice extension
along the $x$ direction (in lattice units) and 
the right hand side (RHS) condition is needed to guarantee smoothness 
of the magnetic field across the periodic boundaries~\cite{wiese,review}; 
according to Eq.~(\ref{four_potential}), 
all other abelian links are set to $1$. Notice that, consistently with 
the required zero net magnetic flux across the lattice torus,
the above $U(1)$ links lead to a constant magnetic field but for a single 
plaquette, which is pierced by an additional Dirac string: invisibility 
of that string leads to a quantization condition for the magnetic field~\cite{thooft,bound3,wiese,review}
\begin{equation}
  q_f B=\frac{ 2 \pi b_z}{a^2L_xL_y} \implies
  e B=\frac{ 6 \pi b_z}{a^2L_xL_y}, \qquad b_z \in \mathbb{Z} \, ,
\end{equation}
considering that the smallest quark charge is $e/3$.
The external field is finally added to the discretization of 
$N_f = 2+1$ QCD decribed above by the following substitution
in the Dirac operator 
in Eq.~(\ref{dirac_operator}):
\begin{equation}
  U^{(2)}_{i;\mu} \to u_{i;\mu}^f U^{(2)}_{i;\mu} \, .
\end{equation}
Notice that in this approach the e.m.~field is treated as purely external,
neglecting the back-reaction of quarks on it, and meaning in practice 
that no additional integration over the $U(1)$ gauge 
links is introduced in the partition function.

Bare masses and gauge coupling values have been set in order to move 
on a line of constant physics, 
determined in Refs.~\cite{tcwup1,befjkkrs,physline3} to reproduce experimental results 
for hadronic observables at zero temperature in the continuum limit. 
The introduction of the external field leads to 
additional, $B$-dependent discretization errors. In particular, one should 
consider that the magnetic field acts in practice through the gauge
invariant $U(1)$ phase factors
that dynamical quarks pick going through 
closed loops on the lattice: the smallest non-trivial 
such loop is the plaquette in the $xy$ plane, for which the phase factor is
\beq
\exp\left(i q_f B a^2\right) = \exp\left(i \frac{6 \pi b_z}{L_x L_y} \frac{q_f}{e} \right) \, .
\eeq
Systematic errors in the discretization of the magnetic field 
are under control if such phase is much smaller than $2 \pi$: for the 
up quark, which has the largest electric charge
$q_u = 2 e /3$, the condition reads:
\beq
\frac{2 b_z}{L_x L_y} \ll 1 \, ;
\label{uvemcondition}
\eeq
a useful way to visualize such systematics is to think that
the existence of this minimal phase pickable by dynamical up quarks
is like saying that we are approximating a circle by a regular polygon
with $\sim$ $L_x L_y / (2 b_z)$ sides. All that also sets a natural
UV cut-off for the largest magnetic fields which are explorable 
for a given lattice spacing, which is roughly 
$e B \leq 2 \pi / a^2$.
We have spent a few additional
words on these aspects, since this will be essential
to properly discuss discretization effects in our investigation,
where extremely strong magnetic fields (one order of magnitude
larger than the standard QCD scale) are considered.

\subsection{Observables}

To get the static potential of a $q\bar{q}$-pair, 
similarly to Refs.~\cite{strongmag0,strongmag1}, 
we studied the Wilson loop $\left< \tr W(a\vec{n},an_t) \right>$
and its dependence on the Euclidean time $a n_t$, 
exploiting the relation
\begin{equation}
  \left< \tr W(a\vec{n},an_t) \right> \propto e^{-aV(a\vec{n})n_t},
\end{equation}
which holds for large enough $an_t$. In particular,
from previous equation one can derive
\begin{equation}\label{operative_aV}
  aV(a\vec{n})=\lim_{n_t\to\infty}\log\left({\frac{\left< \tr W(a\vec{n},an_t) \right>}{\left< \tr W(a\vec{n},a(n_t+1)) \right>}}\right) \, ,
\end{equation}
so that 
the potential at fixed $\vec{n}$ can be obtained by fitting to a constant the $\log$ in the RHS of Eq.~(\ref{operative_aV}) as a function
of $n_t$, at least in a suitable stability range.

The color flux tube instead was studied, following Ref.~\cite{tusso}, 
by means of the connected correlator-probe scheme~\cite{Cea:2017ocq,DiGiacomo,DiGiacomo:1990hc,Cea:1992vx,Cea:2012qw,Cea:2014uja,Baker:2019}: 
the observable computed to derive chromoelectric field in-between
the static quark-antiquark pair is in this case:
\begin{multline}\label{flux_op}
  \rho^{\mu t}_{conn}(x_t)=
\frac{\left<\tr (W(an_\mu,an_t)LP^{\mu t}(x_t)L^\dagger)\right>}{\left<\tr (W)\right>}\\ - \frac{\left< \tr (W(an_\mu,an_t))\tr(P^{\mu t}(x_t))\right>}{3\left< \tr (W)\right>},
\end{multline}
where $W$ is the open Wilson loop, $P^{\mu t}$ is the open plaquette 
in the $\mu t$-plane and $L$ the Schwinger path linking the former two 
operators, while $x_t$ is the distance between the plaquette and 
Wilson loop plane, i.e.~the distance from the quark-antiquark axis.
One can easily prove that, in the naive continuum limit,
\begin{equation}\label{eq:conn_cont}
\rho_{conn}^{\mu t}\simeq a^2 g_0\frac{\langle\tr[iWLF_{\mu t}L^{\dag}]\rangle}{\langle \tr(W)\rangle}\ ,
\end{equation}
which can be considered as a probe of the color field strength 
induced by the presence of the quark-antiquark pair, i.e., with 
some abuse of notation, as $a^2g_0\langle F_{\mu t}\rangle_{Q\bar{Q}}$.

\begin{figure}[t!]
\includegraphics*[width=0.9\columnwidth]{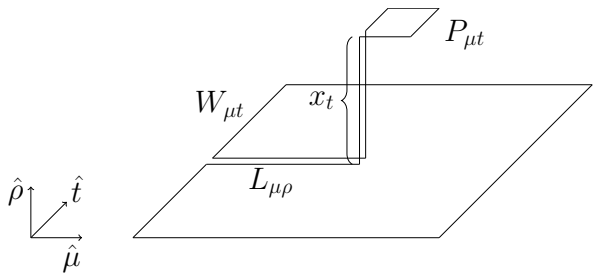}
\caption{The representation of the $SU(3)$ path $\rho^{\mu t}_{conn}$ defined in Eq.~(\ref{flux_op}). Since we are interested in the longitudinal 
chromoelectric field only, $P$ and $W$ always lie on parallel planes.}
\label{fig_operator} 
\end{figure}

The connected correlator $\rho_{conn}$ is pictorially described in
Fig.~\ref{fig_operator}: $L$ is attached to the square
Wilson loop $W$ in the midpoint of its temporal extent, 
it reaches half the
distance between the quark-antiquark pair and then it moves $x_t$ lattice
spacings in one of the directions orthogonal to the plane of the Wilson loop:
in this way the flux tube profile is determined at the midpoint of the static
color source.
In this case, as in Ref.~\cite{tusso},
the investigation has been limited to squared Wilson loops.

It is important to note that the presence of the background magnetic field along $\hat{z}$ breaks the spatial octahedral symmetry, leaving a $D_4$ symmetry on the $xy$-plane. In the evaluation of the static potential by 
means of the Wilson loop, this condition implies that loops in the $z$-direction are not equivalent to those extending in $x$- and $y$-directions, which on the other hand are equivalent to each other: 
that naturally leads to distinguish between a static potential measured longitudinally (L)
to the magnetic field, or transverse (T) to it. Actually, one can consider also generic angles 
between the magnetic field direction and the quark-antiquark axis: this is best done
by considering magnetic fields with a generic orientation relative to the lattice axes.
This kind of more general analysis has been performed in Ref.~\cite{strongmag1}, showing however 
that most of the angular dependence of the static potential can be accounted for by the lowest harmonic.
That means that the relevant information is contained in the L and T-potentials, which are therefore
the only cases considered in the present investigation.

The study of the correlator $\rho^{\mu t}_{conn}$ is a bit more involved, due to its three-dimensional shape.
Apart from the T- or L-cases characterizing the orientation of the quark-antiquark separation relative 
to the magnetic field, in the T-case one can further distinguish whether the flux tube
profile is studied in the direction parallel or orthogonal to $\vec B$: the analysis of 
Ref.~\cite{tusso} shows that some minor anisotropies emerge also in this case, i.e.~the flux tube
itself loses its axial symmetry. Therefore, as for the flux tube profile, we will consider
three different cases: L, TL and TT. {We denote by $\hat{\mu}$ the direction of the quark-antiquark axis, so that the possible geometries can be mapped into these three equivalence classes according to Table~\ref{table:classes}.}
\begin{table}
	\centering
	\begin{tabular}{|c|c|c|}
		\hline 
		\parbox{1cm}{\centering $\hat{\mu}$}  & \parbox{1cm}{\centering $\hat{\rho}$} & class \\ \hline 
		\hline
		$\hat{z}$                  & $\hat{x}$                  & \multirow{2}{*}{L} \\ 
		$\hat{z}$                  & $\hat{y}$                  & \\ \hline
		$\hat{x}$                  & $\hat{y}$                  & \multirow{2}{*}{TT} \\ 
		$\hat{y}$                  & $\hat{x}$                  & \\ \hline
		$\hat{x}$                  & $\hat{z}$                  & \multirow{2}{*}{TL} \\ 
		$\hat{y}$                  & $\hat{z}$                  & \\ \hline
		
	\end{tabular}
	\caption{{Equivalence classes of the relative orientations between the quark-antiquark pair, the magnetic field (fixed along the $\hat{z}$ direction) and the transverse direction $x_t$. Labels $\hat{\mu}$ and $\hat{\rho}$ refer to Fig.~\ref{fig_operator}. Letters T and L stand for transverse or longitudinal with regard to the magnetic field.}}
	\label{table:classes}
\end{table}%
{A residual symmetry $x_t \to - x_t$ is preserved but, anyway, it has not been exploited in this work.}

For the evaluation of both observables
in order to reduce the UV noise, we applied 
one step of
HYP smearing~\cite{Hasenfratz:2001hp} for 
temporal links, 
with the following choice of parameters: $\alpha_1=\alpha_2=2\alpha_3=1$ (as for the HYP2-action defined in Ref.~\cite{Della Morte:2005yc}). Moreover, we performed several steps of (spatial) APE smearing~\cite{Albanese:1987ds} on the spatial links, so that the $N$-times smeared link reads
\begin{equation}
 U_{i;\mu}^{(N)}=\left[{U_{i;\mu}^{(N-1)}+\alpha_{APE}S_{i;\mu}^{(N-1)}}\right]_{SU(3)},
\end{equation}
where $U_{i;\mu}^{(0)}=U_{i;\mu}$, $S_{i;\mu}^{(N-1)}$ is the sum of the spatial staples around the link $U_{i;\mu}^{(N-1)}$, $[\cdot]_{SU(3)}$ denotes the projection on the gauge group and 
{the choices for $\alpha_{APE}$ match those of previous works: $0.25$ for the string tension as in~\cite{strongmag1} and $1/6$ for the QCD flux tube as in~\cite{tusso}.}


\section{Numerical Results}
\label{results}

\begin{table}[b!]
  \centering
  \begin{tabular}{ c c c c c }
    \hline
    lattice size    &   $a [\textrm{fm}]$ & $\beta$   & $am_s$ & $b_z$ \\ \hline
    $24^3\times48$  & 0.114       & 3.787      & 0.0457 & 0{,}41{,}93   \\   
    $32^3\times64$  & 0.086      & 3.918      & 0.0343 & 0{,}41{,}93   \\   
    $48^3\times96$  & 0.057      & 4.140      & 0.0224 & 0{,}41{,}93   \\
    \hline
  \end{tabular}
  \caption{Simulation parameters based on~\cite{tcwup1,befjkkrs,physline3} and corresponding to physical values of the pion mass. The strange-to-light mass ratio is $m_{s}/m_{u,d}=28.15$. 
The
systematic error on $a$ is about $2 - 3~\%$~\cite{tcwup1,befjkkrs,physline3} .}
    \label{table:parametri}
\end{table}

Our results are based on simulations corresponding
to three different values of $e B$ (0, 4 and 9 GeV$^2$) and three
different lattice spacings in each case ($a \simeq 0.057, 0.086$ and
 0.114~fm) in order to allow for a continuum extrapolation;
simulations at zero magnetic field have been performed
mostly for renormalization purposes.
The spatial lattice size has been kept fixed in most cases to $a L_s \sim 2.75$~fm, with an Euclidean temporal extent
twice as large.
A summary of all simulation
points is reported in Table~\ref{table:parametri}.
Monte-Carlo sampling of gauge configurations has been
performed based on a Rational Hybrid Monte-Carlo (RHMC) algorithm running on 
GPUs~\cite{openacc1,openacc2}. For each simulation we performed $O(10^3)$ RHMC steps, 
taking measures every $10$ unit trajectories. 
The statistical analysis has been based in most cases 
on a binned bootstrap analysis.

\begin{figure}[t!]
\includegraphics*[width=0.9\columnwidth]{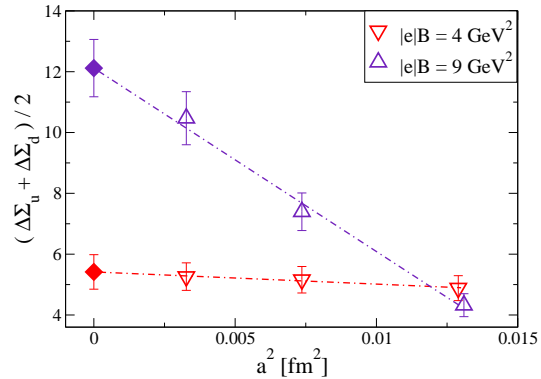}
\caption{Change of the renormalized average light quark condensate due 
to the magnetic field for $eB = 4$ and 9~GeV$^2$ as a function of the 
lattice spacing, together with continuum extrapolations obtained
assuming $O(a^2)$ corrections.}
\label{cond_continuum} 
\end{figure}

In our analysis we will assume the lattice spacing 
being independent of $eB$. This is a reasonable assumption
as long as the magnetic field is much smaller than
the UV cutoff, hence it is expected to lead to sensible
results at least when continuum extrapolations are considered.
As a matter of fact, the assumption has been explicitly 
checked only for smaller values of the magnetic field~\cite{Bali:2011qj};
however our analysis of the chiral condensate, leading to results
in agreement with theoretical expectations, will further 
support the hypothesis.

\subsection{Magnetic catalysis at extremely large magnetic 
fields}

\label{catalysis}

Before focusing on the confining aspects of the theory,
let us discuss its chiral properties, for which predictions
are well established. In particular one expects, at least for 
$T= 0$, the magnetic catalysis phenomenon, with an enhancement of chiral
symmetry breaking induced by the magnetic background field and 
detectable as an increase of the chiral condensate.

\begin{figure}[t!]
\includegraphics*[width=0.9\columnwidth]{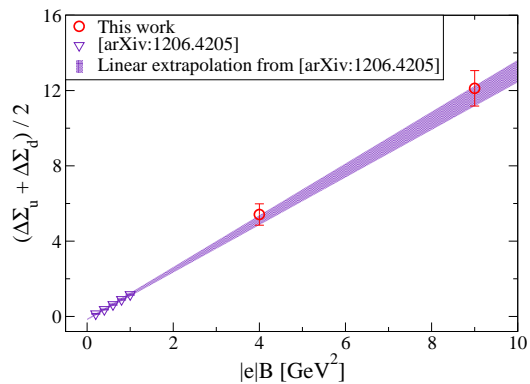}
\caption{Continuum extrapolated results for the change of the chiral condensate
due to the magnetic background field: results obtained in this study are compared
to those reported in Ref.~\cite{Bali:2012cd}. The colored band is the result of a linear
fit to the data of Ref.~\cite{Bali:2012cd} ($\chi^2/d.o.f. \simeq 1/3$), which after
extrapolation turns out to be in nice agreement, within errors, 
with our present determinations.}
\label{cond_compare} 
\end{figure}

In order to compare with previous results in the literature, we will
consider the change of the light quark condensate 
due to the magnetic field, renormalized as 
in Ref.~\cite{Bali:2012cd}:
\beq
\Delta \Sigma_q(B)= \frac{2 m_q}{m_\pi^2 F_\pi^2} 
(\langle \bar \psi \psi_q \rangle_B - 
\langle \bar \psi \psi_q \rangle_{B = 0})
\label{def_condensate}
\eeq
with $q = u,d$, where 
$\langle \bar \psi \psi_q \rangle$ is determined as usual 
in terms of the volume normalized trace of the inverse fermion
matrix (computed by noisy estimators),
$m_\pi = 135$~MeV is the pion mass and 
$F_\pi= 86$~MeV is the pion decay constant in the chiral limit.

The average quantity 
$(\Delta \Sigma_u + \Delta \Sigma_d)/2$ is displayed 
in Fig.~\ref{cond_continuum} for the two explored values
of $eB$ as a function of the squared lattice spacing $a^2$, together
with a continuum extrapolation obtained assuming 
$O(a^2)$ corrections. It is interesting to notice that continuum
corrections are significantly larger for $eB = 9$~GeV$^2$: that can
be easily understood in terms of what discussed above regarding 
the $B$-dependent discretization errors (see Eq.~(\ref{uvemcondition})
and comments thereafter). For $eB = 4$~GeV$^2$, corresponding to $b_z = 41$,
these kind of discretization errors, at the three different lattice 
spacings, can be put in analogy with those 
that one has by approximating a circle by a regular polygon with
respectively (from coarsest to finest) 7, 12, 28 sides, which is reasonable right
from the beginning; for $eB = 9$~GeV$^2$ instead, corresponding to $b_z = 93$,
the approximation starts with a triangle (which is far from good) and ends with a dodecagon 
(which is reasonable). These considerations make it clearer why, having in mind to perform
a reliable continuum extrapolation, it is not reasonable to consider larger values of $eB$,
unless smaller lattice spacings are computationally affordable.

In Fig.~\ref{cond_compare} continuum extrapolated results are 
compared to the analogous ones obtained, for $eB \leq 1$~GeV$^2$,
in Ref.~\cite{Bali:2012cd}. The large field behavior of the magnetic
catalysis phenomenon can be thoretically predicted in terms
of a lowest Landau level (LLL) approximation.
The higher energy levels increase proportionally to 
$\sqrt{eB}$, thus they become practically 
irrelevant to the dynamics of the system. 
The LLL is instead independent of $eB$, 
while its degeneracy linearly increases with it: that 
leads to predict a linear behavior in the density
of near-zero modes, hence in the chiral condensate by
the Banks-Casher relation. This linear behavior is nicely
reproduced in Fig.~\ref{cond_compare}: in particular,
in the figure we display the result of a linear fit 
to data from Ref.~\cite{Bali:2012cd} which, when extrapolated 
to the large magnetic fields explored in this study, is perfectly
compatible with our results within errors.

Notice that the lattice
spacing enters with a fourth power in fixing 
the renormalization group invariant quantity 
in Eq.~(\ref{def_condensate}): hence, the nice consistency with 
data from Ref.~\cite{Bali:2012cd} and with
the LLL prediction supports the assumption 
that the lattice spacing is indeed independent of $B$.

\subsection{Static quark-antiquark potential}

\begin{figure}[t!]
\includegraphics*[width=0.9\columnwidth]{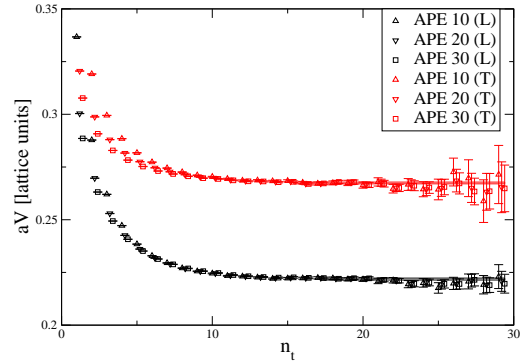}
\caption{{Logarithm of Wilson loop ratio according to Eq.~(\ref{operative_aV}) as a function of $n_t$. Data have been extracted from the $48^3\times96$ lattice ($a=0.0572$~fm) at $eB = 9$~GeV$^2$. They are displayed for three choices of the APE smearing level in both the T and L cases. Continuum lines correspond to the determination of the plateau for $N_{APE}=30$}. }
\label{stab_smear} 
\end{figure}

The static quark-antiquark potential has been derived as described above,
by looking for a plateau, as a function of the temporal extent $n_t$,
for the logarithm of Wilson loop ratios reported 
in Eq.~(\ref{operative_aV}). 
An example is showed in Fig.~\ref{stab_smear}, where we report data
obtained for both the trasverse and longitudinal direction at
$eB = 9$~GeV$^2$ and $R = 5 $ for the finest lattice spacing.
The two bands show our final determination of the 
potential for the two cases and have been obtained considering
Wilson loops after 30 spatial APE smearing steps, however we report
in the figure also data obtained after 10 and 20 smearing steps, which
are practically indistinguishable. A similar stability under APE smearing
is observed for all explored values of $e B$, lattice spacing and
quark-antiquark separation $R$.

\begin{figure}[t!]
	\includegraphics*[width=0.9\columnwidth]{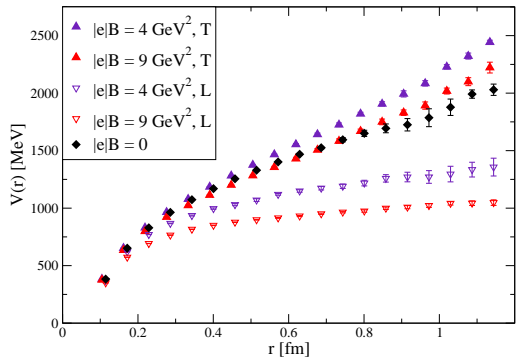}
	\caption{{Static potential $V(r)$ between the quark-antiquark pair as a function of the distance $r$, for the explored values and orientations
of $\vec{B}$. Results refer to the finest lattice spacing $a=0.0572$~fm, i.e. the $48^3\times96$ lattice.}  }
	\label{potential_finest} 
\end{figure}

In Fig.~\ref{potential_finest} we show the final determination of the 
static potential obtained for the finest lattice spacing and
all the explored values of $eB$. The anisotropy which is present when 
$eB \neq 0$ is clearly evident, even if in the transverse direction
it shows a non-monotonic behavior with $eB$, with a tendency for a 
slight decrease of the slope when going from 4 to 9~GeV$^2$,
at least for this value of the lattice spacing.

\begin{figure}[t!]
\includegraphics*[width=0.9\columnwidth]{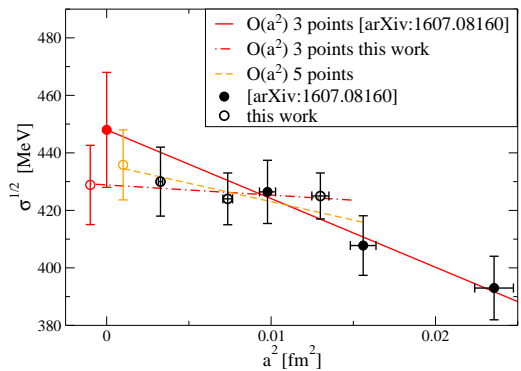}
\caption{Continuum limit of the string tension at $eB = 0$, together with the results of Ref.~\cite{strongmag1}.}
\label{compare_sigma_eBzero} 
\end{figure}

As a preliminary analysis, we have considered results for 
the potential at $eB = 0$ and compared them with previous results
in the literature, in order to check consistency. In particular,
in Fig.~\ref{compare_sigma_eBzero} we compare results obtained 
for the string tension in this work with those obtained in Ref.~\cite{strongmag1}
using the same lattice discretization but different lattice spacings.
The two sets of results are perfectly compatible with each other 
and a combined continuum extrapolation assuming $O(a^2)$ corrections
returns a continuum value $\sqrt{\sigma} = 435(8)$~MeV with 
$\chi^2 / {\rm d.o.f.} = 1.3/3$ (the point on the coarsest lattice being 
discarded), which is perfectly compatible with phenomenological 
predictions and lattice determinations for $\sigma$~\cite{Aoki:2016frl}. We would like to stress the importance 
of this consistency check: since the physical spatial lattice sizes adopted 
in this work and in Ref.~\cite{strongmag1} are, for computational reasons,
quite different ($\sim 3$~fm vs $\sim 5$~fm), the agreement we find shows
that, at least for what concerns the static quark-antiquark potential, finite size effects are not
significant.

\begin{figure}[t!]
	\includegraphics*[width=0.9\columnwidth]{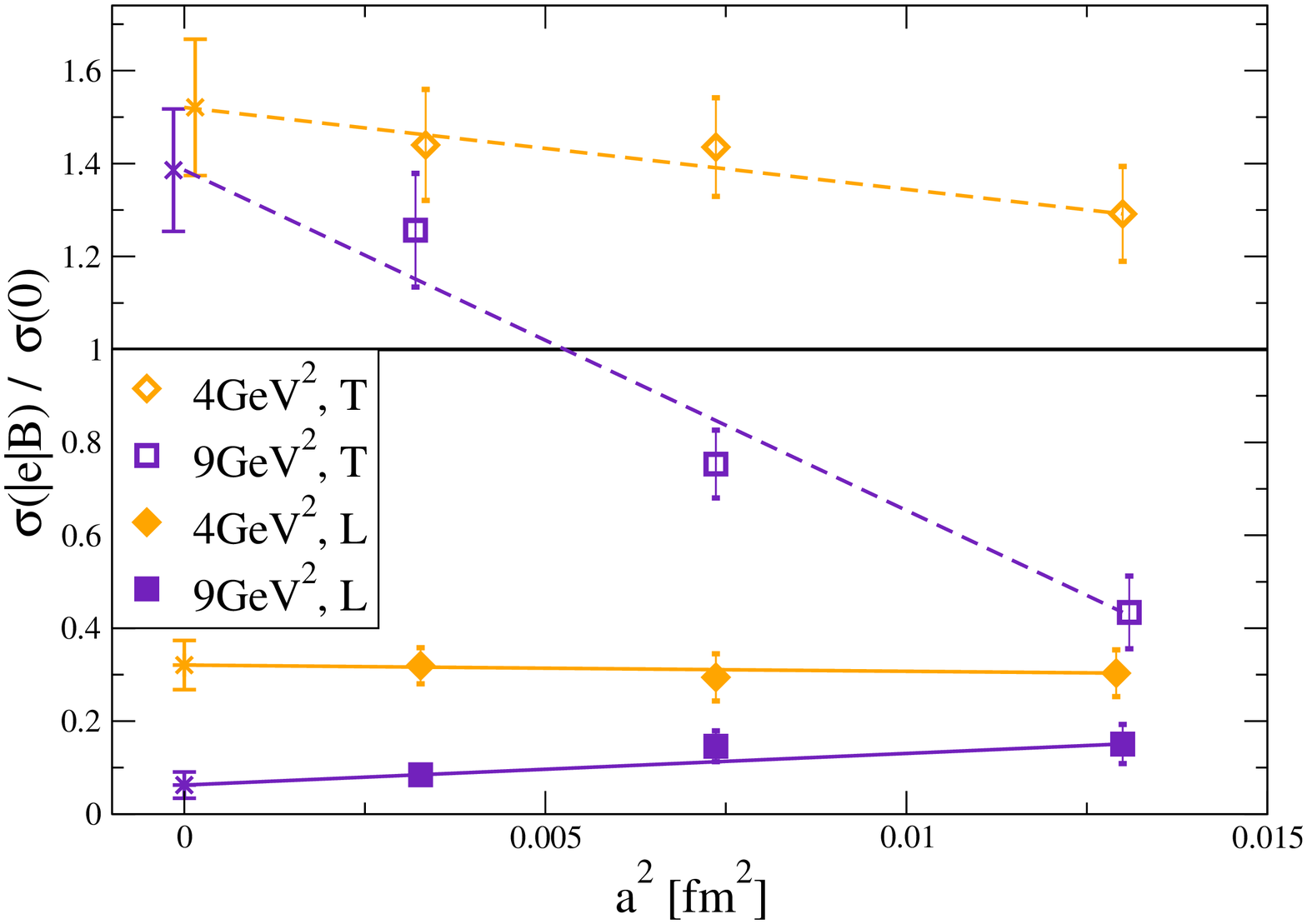}
	\caption{{Continuum limit of the $\sigma$-ratios for both the values of $B$. Dashed/continuum lines correspond to the best extrapolations performed in T/L cases.}}
	\label{sigma_ratios} 
\end{figure}

Next, in order to assess to the fate of the static potential
anisotropy in the explored range of magnetic fields, we consider,
as in Ref.~\cite{strongmag1}, the dimensionless ratios 
$\sigma(eB) / \sigma(0)$, which are reported
in Fig.~\ref{sigma_ratios}
 as a function of $a^2$ for both magnetic fields and for
both the longitudinal and the transverse directions, together
with continuum extrapolations assuming $O(a^2)$ corrections.
Regarding the string tension in the longitudinal direction,
we confirm the findings of Ref.~\cite{strongmag1}: it is a consistently decreasing function
of $eB$, both for at finite lattice spacing and in the continuum limit;
however, contrary to the hypothesis put forward in Ref.~\cite{strongmag1},
we find a non-zero string tension, within two standard deviations, even at 
the largest explored value of $e B$. Regarding the transverse direction, instead, 
we can appreciate from Fig.~\ref{sigma_ratios} that, at least for finite lattice spacing,
the trend for an increasing string tension observed in Ref.~\cite{strongmag1}
seems inverted. However, when considering continuum extrapolations, one realizes that 
$\sigma(eB) / \sigma(0)$ actually reaches a saturation at large $eB$.

\begin{figure}[t!!]
	\includegraphics*[width=0.9\columnwidth]{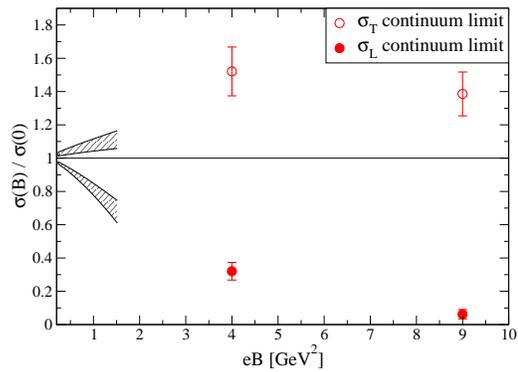}
	\caption{{Continuum limit of the $\sigma$-ratios in the T and L cases, for both the values of the background field $eB=4,9$~GeV$^2$. The dashed gray regions correspond to the continuum extrapolations of Ref.~\cite{strongmag1}}}
	\label{ratios_continuum} 
\end{figure}

Such results are better appreciable in Fig.~\ref{ratios_continuum},
where the continuum extrapolated values for $\sigma(B)/\sigma(0)$ 
in the T- and L-directions obtained in this study are compared
with the continuum extrapolation of Ref.~\cite{strongmag1}, which is 
plotted only in the relevant range of $eB$ where simulations of 
Ref.~\cite{strongmag1} were performed.
Present results are not inconsistent with those of Ref.~\cite{strongmag1},
however they clarify the perspective for the large-$eB$ limit of the 
string tension. In the trasverse direction, the string tension seems to reach
a saturation at a value which is around 50~\% larger than the zero field
value. In the longitudinal direction the string tension keeps decreasing as
a function of $eB$, but is still significantly different from zero for 
$eB \sim 4$~GeV$^2$, contrary to what the continuum extrapolation
of Ref.~\cite{strongmag1} could have suggested: the large field behavior is difficult
to predict precisely, it could be either an exponential-like decreasing behavior 
for which the string tension never vanishes, or a more straight decrease where 
$\sigma_L$ finally vanishing at some critical value for $eB \gtrsim 
10$~GeV$^2$.
Such possibility should be further explored by future studies,
capable of approaching even smaller values of the lattice spacing, which 
has been the main constraint limiting our simulations to $eB \lesssim 10$~GeV$^2$
in order to access properly extrapolated continuum results.

\subsection{Color flux tubes}

\begin{figure}[t!!]
	\includegraphics*[width=0.9\columnwidth]{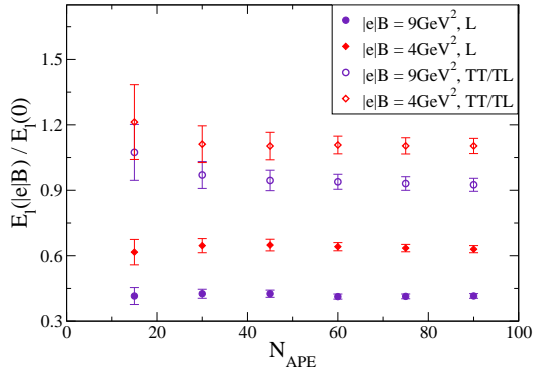}
	\caption{Ratio $E_l(B)/E_l(B=0)$ at $a=0.0572$ fm for different numbers of smearing steps. Data are evaluated at $x_t=0$, so that the transverse configurations TT and TL are equivalent (see  Table~\ref{table:classes} for details). The physical distance between the color charges is $d=0.68$~fm.}
	\label{fig:indep_Nape} 
\end{figure}
The analysis of color flux tubes is expected, in general, to confirm 
results obtained by the analysis of the static quark-antiquark potential:
this is indeed the outcome of Ref.~\cite{tusso}, showing that the main
effect of the magnetic field is an overall suppression/enhancement of 
the flux tube in the L/T directions, with a slight 
modification of its profile, which however can be still 
nicely described by 
models inspired to dual superconductivity of the QCD vacuum.

We measured the longitudinal component $E_l$ of the chromo-electric field (directed along the quark-antiquark axis) since previous studies showed that it is by far the dominant one (see~\cite{Baker:2019} and references therein). We denote by $\hat{\mu}$ the direction where the color charges lay, so that the longitudinal chromo-electric field is given by
\beq
    E_{l}(d,x_t)=\frac{1}{a^2g_0}\rho^{\mu t}_{conn}(d,x_t) \, ,
\eeq
where $d$ is the separation distance between the quark-antiquark pair and $x_t$ is the transverse distance at which the field is probed (see Fig.~\ref{fig_operator}).

The use of smearing techniques introduces a non trivial dependence on the amount of smearing adopted.
On the other hand, the analysis of Ref.~\cite{tusso} showed that ratios of observables with and without the magnetic field, such as $E_l(B)/E_l(B=0)$, are insensitive to the number of smearing steps. We verified that this feature holds true for the extreme magnetic fields investigated in this study, as illustrated in Fig.~\ref{fig:indep_Nape}, where we show the ratios of the chromo-electric fields obtained at $a=0.0572$~fm for each magnetic field choice and inequivalent class of Table~\ref{table:classes}. We display the flux tubes at $x_t=0$ but, anyway, the independence is observed for each value of the transverse distance.
Since we are interested in similar ratios of observables, we unambigously decided to fix $N_{APE}=30$ for the analysis carried out for each lattice spacing and value of $x_t$, so that the dependence on $N_{APE}$ will be dropped in the following.

 \begin{figure}[t!!]
 	\includegraphics*[width=0.9\columnwidth]{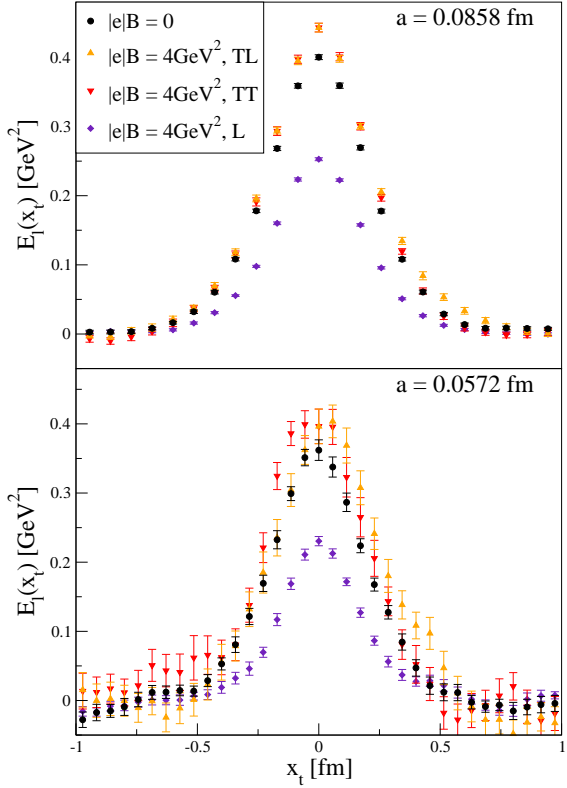}
 	\caption{Chromo-electric field at $eB=4$ GeV$^2$ for two choices of the lattice spacings $a=0.0858,0.0572$ fm, in the physical range $x_t \in [-1,+1]$fm. The relative distance of the quark-antiquark pair is fixed to $d=0.68$ fm.}
 	\label{fig:tubes4} 
 \end{figure}
 \begin{figure}[t!!]
 	\includegraphics*[width=0.9\columnwidth]{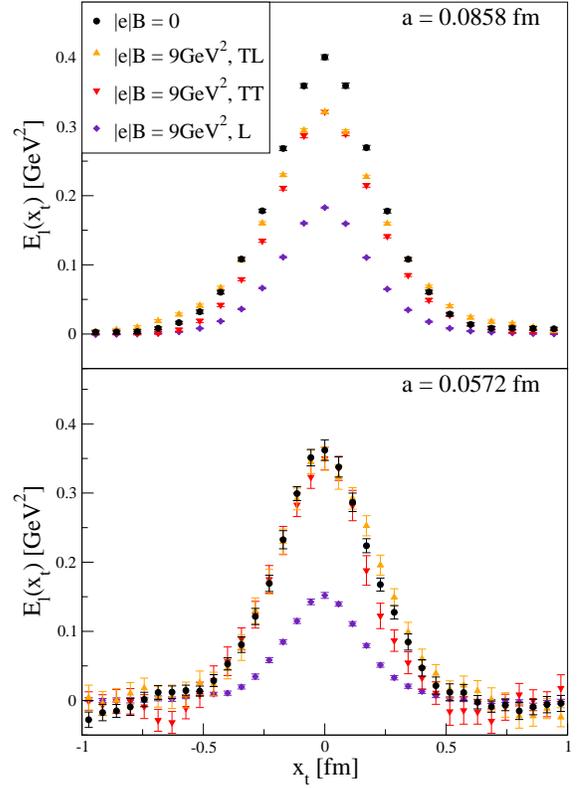}
 	\caption{Chromo-electric field at $eB=9$ GeV$^2$ for two choices of the lattice spacings $a=0.0858,0.0572$ fm, in the physical range $x_t \in [-1,+1]$fm. The relative distance of the quark-antiquark pair is fixed to $d=0.68$ fm.}
 	\label{fig:tubes9} 
 \end{figure}
In Fig.~\ref{fig:tubes4} and Fig.~\ref{fig:tubes9} we show some results for the flux tube extracted from simulations performed at lattice spacings $a=0.0858, 0.0572$~fm, using Wilson loops of spatial size $d=0.68$~fm.
The influence of the background field on the color flux tube is compatible with the findings of the previous section: strong anisotropies are induced depending on the magnitude and orientation of the external field. In detail, in the L case the flux tube monotonically decreases as the magnetic field grows; in transverse cases (TT-TL), the chromo-electric field is enhanced at $eB=4$~GeV$^2$ while a non-trivial dependence on the lattice spacing is exhibited for $eB=9$~GeV$^2$: the flux tube is suppressed by the magnetic field at $a=0.0858$~fm and compatible with the $eB=0$ case at $a=0.0572$~fm. We stress that this trend is consistent with the scaling dependence on $a$ observed for the string tension extracted in transverse cases at $eB=9$~GeV$^2$ (see Fig.~\ref{sigma_ratios}).

\begin{figure}[t!!]
	\includegraphics*[width=1\columnwidth]{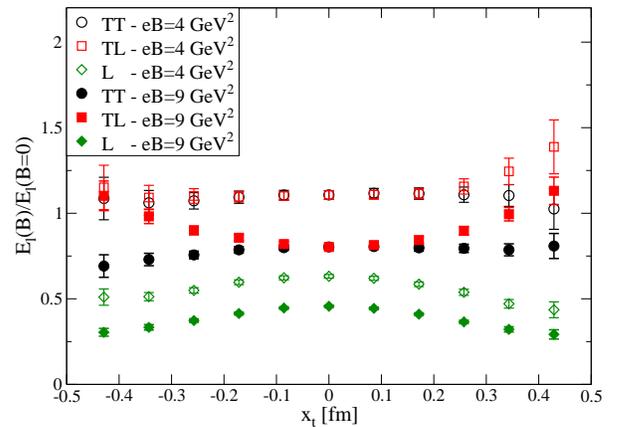}
	\caption{Ratio $E_l(B,x_t)/E_l(0,x_t)$ as function of the transverse distance $x_t$. Data have been computed for $eB=4$ and $9$~GeV$^2$ and for each orientation class at $a=0.0858$~fm. The physical distance between the pair is fixed to $d=0.68$~fm.}
	\label{fig:ratioE} 
\end{figure}

\begin{figure}[t!!]
	\includegraphics*[width=1\columnwidth]{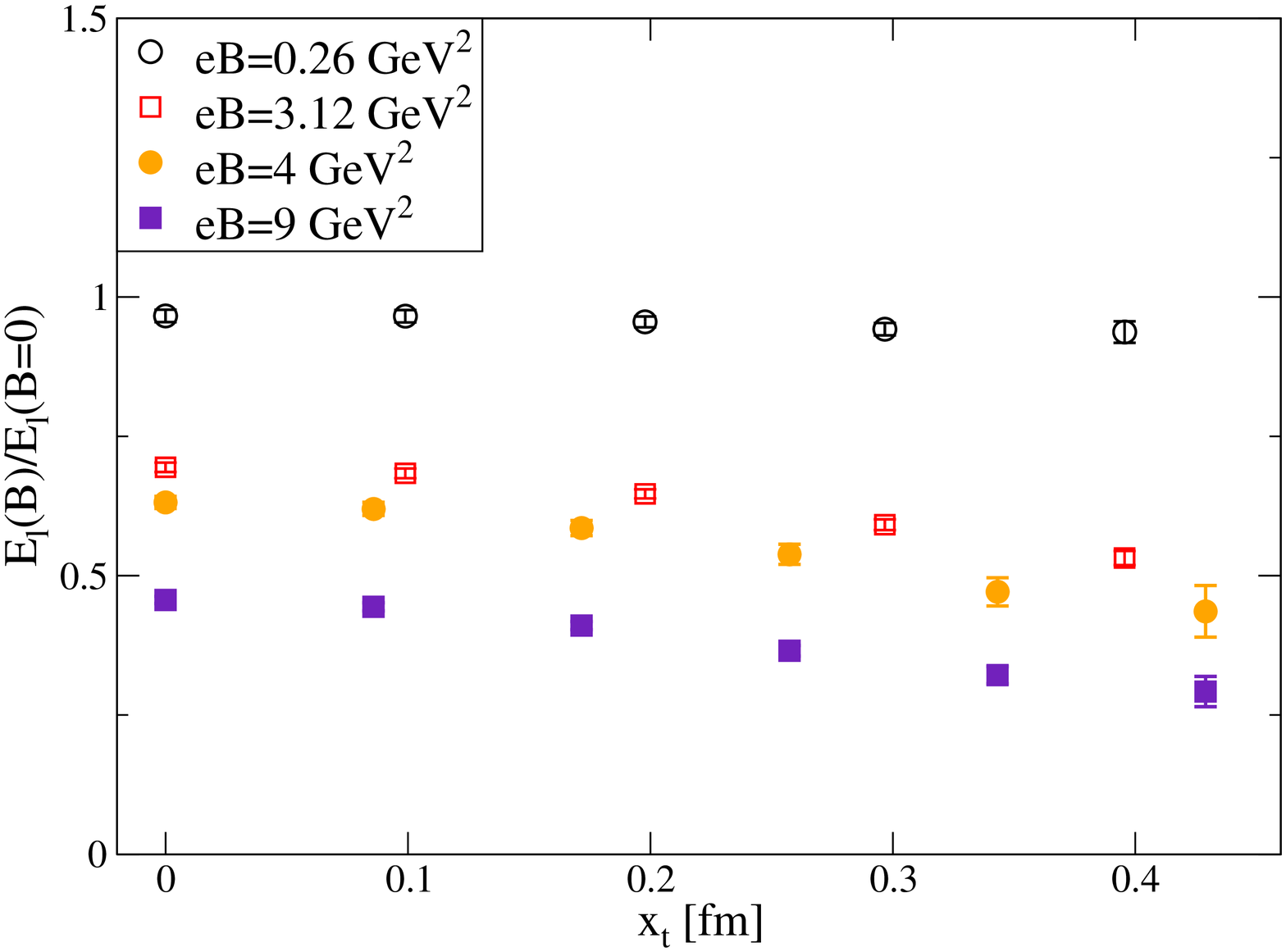}
	\caption{Ratio $E_l(B,x_t)/E_l(0,x_t)$ computed in L configurations. Data for $eB=4$ and $9$~GeV$^2$ have been computed at $a=0.0858$~fm and $d=0.68$~fm. Values at weaker fields were computed in Ref.~\cite{tusso} at $a=0.0989$~fm and $d\simeq0.7$~fm.}
	\label{fig:compare_tusso} 
\end{figure}

In Fig.~\ref{fig:ratioE} we show the ratio $E_l(B,x_t)/E_l(0,x_t)$ for each magnetic field value and geometry class at $a=0.0858$~fm. Results clearly point out the loss of the cylindrical symmetry in transverse configurations, since the TT an TL cases are not equivalent. Furthermore, both at $eB=4$ and $9$~GeV$^2$ the ratio $E_l(B,x_t)/E_l(0,x_t)$ is a decreasing function of $x_t$ in the L case, meaning that the color flux tube gets squeezed by the background field. This behaviour had already been outlined at weaker fields in Ref.~\cite{tusso}, whose results are displayed together with our findings in Fig.~\ref{fig:compare_tusso}.
The comparison points out two effects:
\begin{itemize}
	\item the squeezing phenomenon seems to reach a saturation for large fields, since the decreasing dependence on $x_t$ is quite similar for $eB\ge3.12$~GeV$^2$;
	\item flux tubes are monotonically suppressed by the increasing background magnetic field. On the other hand, a qualitative weakening of the dependece on $eB$ can be noticed, in agreement with the behaviour of the string tension outlined in Fig.~\ref{ratios_continuum} and discussions thereafter.
\end{itemize}

\begin{figure}[t!!]
	\includegraphics*[width=1\columnwidth]{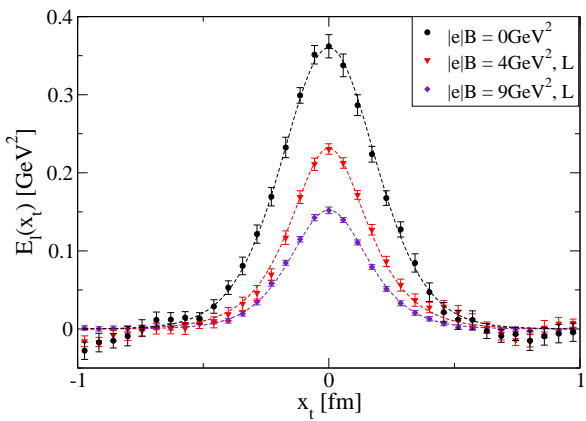}
	\caption{Dependence of the color flux tube profile on the intensity of the magnetic field in L configurations ($a=0.0572$~fm and $d=0.68$~fm). The dashed lines represent the best fits according to the Clem ansatz of Eq.~(\ref{eq:Clem}).}
	\label{fig:clem} 
\end{figure}

Despite the observed deformations, the functional dependence 
of the flux tube profile does not seem to change significantly.
We employ a parametrization inspired by the form of magnetic fields inside vortices in type II superconductors to fit the data for the longitudinal chromo-electric field.
In particular, we follow the  parametrization proposed in \cite{Clem}, where an expression for the magnetic flux tube which solves the Ginzburg-Landau equations is obtained by a variational model for the normalized order parameter of an isolated vortex. This expression, often called Clem ansatz, reads 
\beq\label{eq:Clem}
E_l(x_t)=\frac{\phi}{2\pi}\frac{\mu^2}{\alpha}\frac{K_0( \sqrt{\mu^2x_t^2+\alpha^2})}{K_1(\alpha)} \, ,
\eeq
where $K_n$ are the modified Bessel functions of the second kind of order $n$ while $\alpha,\mu$ and $\phi$ are fit parameters. Previous studies showed that the flux tube profile is well described by the Clem function, also in the presence of an external field \cite{tusso}. Remarkably, we find that the expression in Eq.~(\ref{eq:Clem}) is a suitable model even at the large magnetic fields explored in this work. As an example, in Fig.~\ref{fig:clem} we show the chromo-electric fields obtained at the finest lattice spacing in L configurations together with the best fit functions. We checked that the fit works reasonably well for all the choices of magnetic field, geometry class and lattice spacing.

Flux tubes can be used to compute a more significant parameter: the linear energy density $\epsilon(B)$. Since the transverse components of the chromo-electric field are negligible, the energy density just reads
\beq
\epsilon=\frac{1}{2}\int \textrm{d}^2x_t\,E_l(d,x_t)^2 \, .
\eeq
The integration is performed over the section orthogonal to the quark-antiquark pair axis, requiring to explicitly know the angular dependence of the color flux tubes.  However, the chromo-electric field possess a cylindrical symmetry over the plane when $\vec{B}$ is directed along the charges (L configurations), so that it is independent of the azimuthal angle. In this case, the integration procedure can be pursued based on data extracted along just one direction on the plane. Furthermore, a numerical integration is not needed: assuming the expression in Eq.~(\ref{eq:Clem}), then the known integrals of the modified Bessel functions can be exploited (see, e.g., Eq.~(5.52.1) and Eq.~(5.54.2)
in Ref.\cite{integrali}), leading to
\beq
\epsilon=\frac{\phi^2 \mu^2}{8\pi}\left( 1 - \frac{K_0(\alpha)^2}{K_1(\alpha)^2} \right) \, ,
\eeq
so that the linear energy density can be expressed in terms of best fit parameters. This allows to avoid problems regarding the numerical integration instability and the systematic uncertainties which would arise due to the sharp peaks of the flux tube profile.
\begin{figure}[t!!]
	\includegraphics*[width=1\columnwidth]{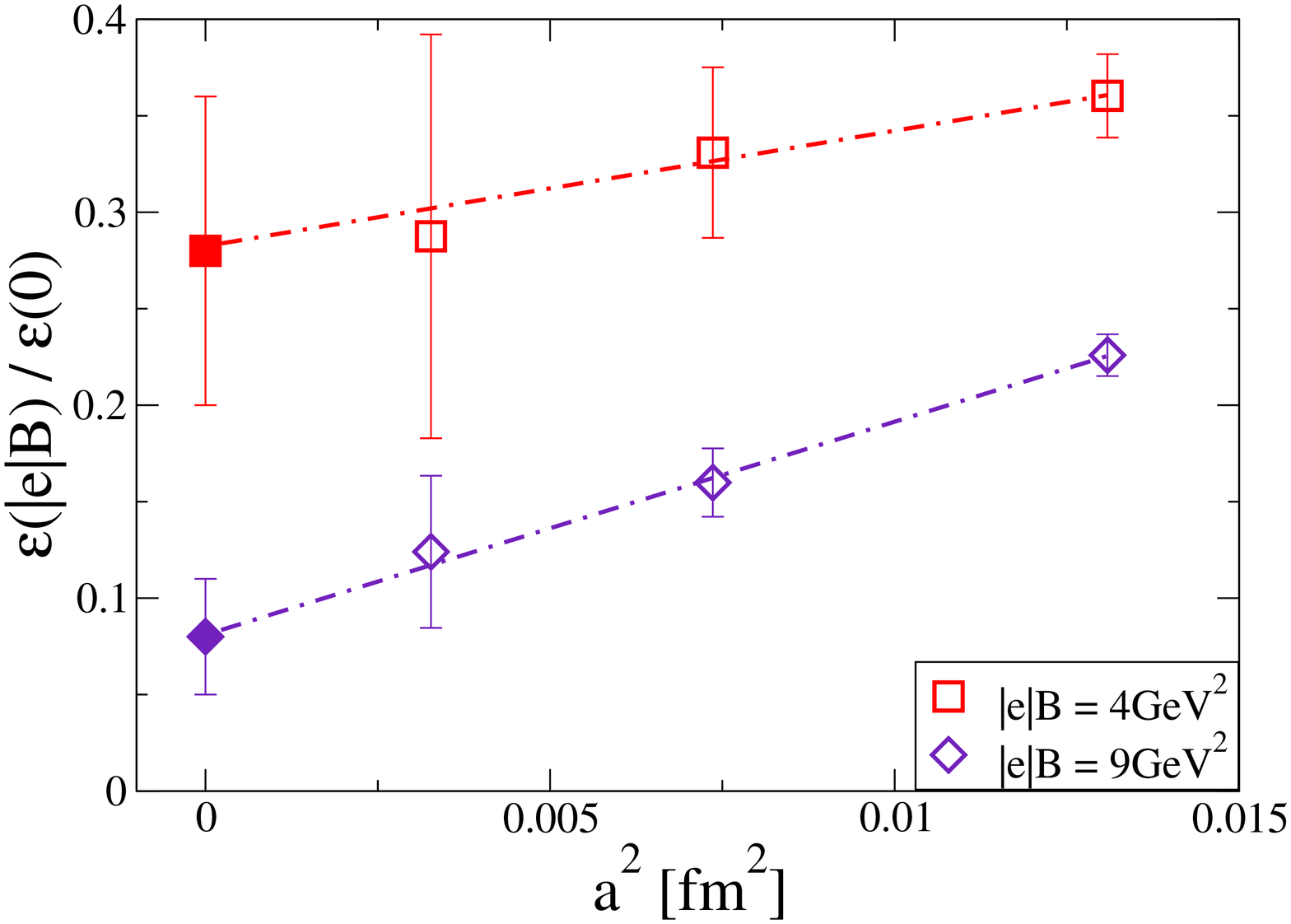}
	\caption{Continuum limits of the ratio $\epsilon(B)/\epsilon(0)$ at $eB=4,9$ GeV$^2$.}
	\label{fig:epslimit} 
\end{figure}
So, we show in Fig.~\ref{fig:epslimit} the ratio $\epsilon(B)/\epsilon(0)$ extracted from L configurations for both the values of the background field and each lattice spacing, together with the continuum extrapolations performed assuming $O(a^2)$ corrections.

\begin{figure}[t!!]
	\includegraphics*[width=1\columnwidth]{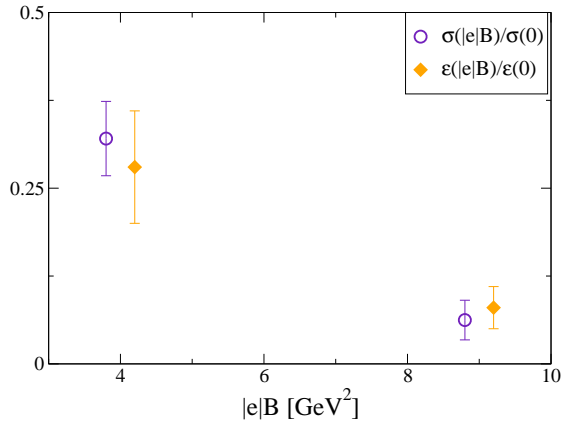}
	\caption{Comparison of $\sigma(B)/\sigma(0)$ and $\epsilon(B)/\epsilon(0)$ computed at $eB=4$ and $9$ GeV$^2$. Points are slightly shifted for readability.}
	\label{fig:compare} 
\end{figure}

In a classical picture, the energy density per unit length is strictly related to the string tension, since the latter is nothing but the slope of the linear term in the potential. A direct comparison is not possible due to the strong dependence of $\epsilon(B)$ on the smearing procedure. However, this issue is overcome by taking into account the ratio $\epsilon(B)/\epsilon(0)$, where the dependence on $N_{APE}$ is expected to disappear. Actually, the fact that the ratio $E_l(B)/E_l(0)$ is independent of the smearing procedure, as seen in Fig.~\ref{fig:indep_Nape}, does not imply \textit{a priori} that ratios of fit parameters (and so $\epsilon(B)$) are independent too. Nevertheless, this independence is numerically observed in the energy density for all the values of $eB$. The comparison is hence possible and it is performed in Fig.~\ref{fig:compare}, where the continuum limit of the ratio $\epsilon(B)/\epsilon(0)$ is shown together with the ratio of the string tension $\sigma(B)/\sigma(0)$ extracted in the longitudinal case: results are in nice agreement, whithin errors.

\section{Conclusions}
\label{conclusions}

The motivation for the present work lies in 
Refs.~\cite{strongmag0, strongmag1, tusso}, and in particular
Ref.~\cite{strongmag1}, where a prediction was made
for a possible vanishing of the string tension for quark-antiquark
separations in the direction 
longitudinal to a magnetic background field and for field values
$eB \gtrsim 4$~GeV$^2$, based however on the extrapolation 
of results obtained from simulations at smaller field values. 
Which kind of new QCD phase could emerge,
if any, where the string tension vanishes in just one direction,
is an intriguing question which deserves an answer.

In order to make progress in this direction, in this 
study we have pushed the range of magnetic background fields
explorable by lattice simulations, with a control over 
the continuum extrapolation, by considering a set of three
different lattice spacings, going 
down to $a \simeq 0.057$~fm, and a discretization of 
$N_f = 2+1$ QCD similar to that of 
Refs.~\cite{strongmag0, strongmag1, tusso}, i.e.~based on stout 
improved rooted staggered fermions. In this way, we 
have been able to reach $eB \simeq 9$~GeV$^2$.

The main result is that, contrary to the expectations of
Ref.~\cite{strongmag1}, the string tension in the 
longitudinal direction is clearly non-vanishing for 
$eB \simeq 4$~GeV$^2$ and still at two standard deviations
from zero even at $eB \simeq 9$~GeV$^2$, where however it 
is suppressed by one order of magnitude with respect 
to its value at zero 
magnetic background.
On the other hand, the enhancement of the string
tension, as a function of $eB$, in the transverse direction 
seems to reach a saturation at around 50~\% of the string tension value at $B = 0$. 

The analysis of the color flux tube 
shows a consistent suppression/enhancement of its overall amplitude, 
with mild modifications of its profile, consistent with those
already observed in Ref.~\cite{tusso}. In particular, one observes 
a mild squeezing of the flux tube of quark-antiquark separations
parallel to the magnetic field, and a loss of 
cylindrical symmetry for transverse separations.
Notwithstanding such deformations, the flux tube profile 
is still describable by models inspired to dual superconductivity 
of the QCD vacuum in all the explored cases.

Finally, the analysis of the chiral condensate shows a persistence 
of magnetic catalysis in the whole range of explored fields,
 with a behavior compatible with a lowest Landau
level approximation, in particular with a linear dependence
of the chiral condensate on $B$ which is in agreement, within
errors, with that already observed for $eB \gtrsim 1$~GeV$^2$
in Ref.~\cite{Bali:2012cd}.

To summarize, present results postpone to even larger magnetic fields
the possibile emergence of a new phase of strong interactions, 
characterized by the vanishing of the string tension for 
quark-antiquark separation parallel to the magnetic field, and 
by other possible associated new phenomena which have 
not been observed so far. The critical field could be not far from
where we are now, since the longitudinal string tension is 
at just two standard deviations from zero at the largest explored field,
however a careful investigation will require simulations on
finer lattices: in the future we plan to put further efforts
along this direction. A different direction is to investigate 
QCD at finite temperature for the same lattice spacings 
and magnetic background fields explored in the present study,
since that could give indications about the phase structure from 
a different perspective: work is in progress along this 
line~\cite{wip}.

\acknowledgments

We thank M. Cardinali for collaboration in the early stages of this study.
Numerical simulations have been performed on the MARCONI and MARCONI100 machines at CINECA, based on the Project IscrB\_STROMAG and on the 
agreement between INFN and CINECA (under projects INF20\_npqcd, INF21\_npqcd).
F.S. is supported by the Italian Ministry of University and Research (MUR) under grant PRIN20172LNEEZ and by INFN under GRANT73/CALAT.

\end{document}